\batchmode
\makeatletter
\def\input@path{{\string"E:/Trabajo Angel/Mis articulos/Fusion 2018/Trajectory PHD filter/\string"}}
\makeatother
\documentclass[british,english]{IEEEtran}
\usepackage[T1]{fontenc}
\usepackage[latin9]{inputenc}
\usepackage{float}
\usepackage{amsmath}
\usepackage{amsthm}
\usepackage{amssymb}
\usepackage{stmaryrd}
\usepackage{graphicx}

\makeatletter

\providecommand{\tabularnewline}{\\}
\floatstyle{ruled}
\newfloat{algorithm}{tbp}{loa}
\providecommand{\algorithmname}{Algorithm}
\floatname{algorithm}{\protect\algorithmname}

\theoremstyle{plain}
\newtheorem{thm}{\protect\theoremname}
\theoremstyle{definition}
\newtheorem{example}[thm]{\protect\examplename}
\theoremstyle{plain}
\newtheorem{prop}[thm]{\protect\propositionname}

\usepackage{cite} 
\usepackage[margin=8pt,font=footnotesize]{caption}
\usepackage{algorithm}
\usepackage{algpseudocode}
\usepackage{amsmath}  % You need this for the math
\allowdisplaybreaks

\makeatother

\usepackage{babel}
\addto\captionsbritish{\renewcommand{\algorithmname}{Algorithm}}
\addto\captionsbritish{\renewcommand{\examplename}{Example}}
\addto\captionsbritish{\renewcommand{\propositionname}{Proposition}}
\addto\captionsbritish{\renewcommand{\theoremname}{Theorem}}
\addto\captionsenglish{\renewcommand{\examplename}{Example}}
\addto\captionsenglish{\renewcommand{\propositionname}{Proposition}}
\addto\captionsenglish{\renewcommand{\theoremname}{Theorem}}
\providecommand{\examplename}{Example}
\providecommand{\propositionname}{Proposition}
\providecommand{\theoremname}{Theorem}

\begin{document}

\title{Trajectory probability hypothesis density filter}

\author{Ángel F. García-Fernández\foreignlanguage{british}{$^{\star}$},
Lennart Svensson\foreignlanguage{british}{$^{\circ}$}\\
\foreignlanguage{british}{{\normalsize{}$^{\star}$}}{\normalsize{}Dept.
of Electrical Engineering and Electronics, University of Liverpool,
United Kingdom}\\
\foreignlanguage{british}{{\normalsize{}$^{\circ}$}}{\normalsize{}Dept.
of Electrical Engineering, Chalmers University of Technology, Sweden}\\
{\normalsize{}Emails: angel.garcia-fernandez@liverpool.ac.uk, lennart.svensson@chalmers.se}}

\maketitle
\thispagestyle{empty}
\begin{abstract}
This paper presents the probability hypothesis density (PHD) filter
for sets of trajectories: the trajectory probability density (TPHD)
filter. The TPHD filter is capable of estimating trajectories in a
principled way without requiring to evaluate all measurement-to-target
association hypotheses. The TPHD filter is based on recursively obtaining
the best Poisson approximation to the multitrajectory filtering density
in the sense of minimising the Kullback-Leibler divergence. We also
propose a Gaussian mixture implementation of the TPHD recursion. Finally,
we include simulation results to show the performance of the proposed
algorithm.
\end{abstract}


\section{Introduction  }

The probability hypothesis density (PHD) filter is a widely used algorithm
for multitarget filtering, which aims to estimate the state of the
targets at the current time, based on random finite sets (RFSs) \cite{Mahler_book14,Vo05,Vo06,Whiteley10}.
The PHD filter fits into the assumed density filtering (ADF) framework
and propagates a Poisson (multitarget) density on the current set
of targets through the prediction and update steps by minimising the
Kullback-Leibler divergence (KLD) \cite{Angel15_d,Mahler_book14}.

The main appealing characteristics of the PHD filter are its low computational
burden and ease of implementation. It avoids the measurement-to-target
association problem and we just need to calculate the PHD of the multitarget
filtering density, which is defined over the single target space.
It also has some drawbacks such as the spooky effect \cite{Mahler_book14}
or the fact that it does not build tracks. The corresponding smoother
\cite{Nadarajah11,Mahler12} does not avoid these problems. Despite
the inability of the PHD filter to provide tracks, track building
procedures have been proposed for some implementations \cite{Lin06,Panta07,Panta09,Lu17}. 

\begin{figure}
\begin{centering}
\includegraphics[scale=0.6]{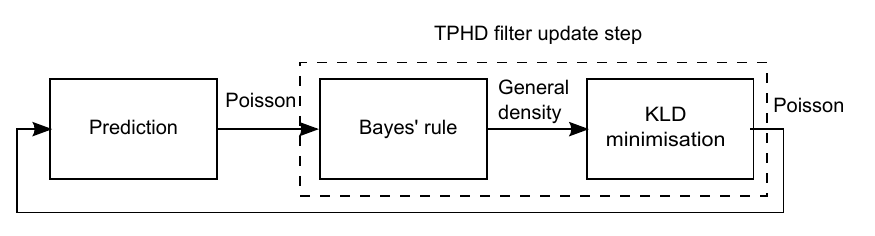}
\par\end{centering}
\caption{\label{fig:TPHD-filter-diagram.}TPHD filter diagram. The TPHD filter
assumes that the multitrajectory densities involved are Poisson (on
the space of sets of trajectories). The output of Bayes' rule is not
Poisson but the TPHD filter obtains the best Poisson approximation
by minimising the KLD.}
\end{figure}

In this paper, we develop a PHD filter that estimates tracks from
first principles: the trajectory PHD (TPHD) filter. The TPHD filter
follows the same scheme as the PHD filter with a fundamental difference,
instead of using a set of targets as the state, it uses a set of trajectories.
The theory for performing multiple target tracking using sets of trajectories
is explained in \cite{Svensson14,Angel15_prov}. A set of trajectories
is a variable that encapsulates the number of trajectories, start
times, lengths and sequence of target states for each trajectory.
In the TPHD filter, we therefore propagate a Poisson (multitrajectory)
density on the space of the set of trajectories through the prediction
and update steps. We do not consider target spawning and assume Poisson
target births so a KLD minimisation is only required after the update
step \cite{Angel15_d}. A diagram of the resulting Bayesian recursion
is given in Figure \ref{fig:TPHD-filter-diagram.}. 

In this paper, we also propose an implementation of the TPHD filter
based on Gaussian mixtures. The resulting Gaussian mixture TPHD (GMTPHD)
filter builds trajectories without the use of labels under a Poisson
approximation whose PHD is represented by a Gaussian mixture. Additionally,
we propose a version of the GMTPHD filter with lower computational
burden called the $L$-scan GMTPHD filter. This filter only updates
the multitrajectory density of the trajectory states of the last $L$
time instant leaving the rest unaltered, which yields an efficient
implementation.

\section{Background\label{sec:Background} }

This section describes some background material on sets of trajectories.
More details can be found in \cite{Angel15_prov}.

\subsection{Variables\label{subsec:Variables}}

A single target state $x\in\mathbb{R}^{n_{x}}$ contains information
of interest about the target, e.g., its position and velocity. A set
of single target states $\mathbf{x}$ belongs to $\mathcal{F}\left(\mathbb{R}^{n_{x}}\right)$
where $\mathcal{F}\left(\mathbb{R}^{n_{x}}\right)$ denotes the set
of all finite subsets of $\mathbb{R}^{n_{x}}$. We are ultimately
interested in estimating all target trajectories, where a trajectory
consists of a sequence of target states that can start at any time
step and end any time later on. Mathematically, a trajectory is represented
as a variable $X=\left(t,x^{1:i}\right)$ where $t$ is the initial
time step of the trajectory, $i$ is its length and $x^{1:i}=\left(x^{1},...,x^{i}\right)$
denotes a sequence of length $i$ that contains the target states
at consecutive time steps of the trajectory. 

We consider trajectories up to the current time step $k$. As a trajectory
$\left(t,x^{1:i}\right)$ exists from time step $t$ to $t+i-1$,
variable $\left(t,i\right)$ belongs to the set $I_{(k)}=\left\{ \left(t,i\right):0\leq t\leq k\,\mathrm{and}\,1\leq i\leq k-t+1\right\} $.
A single trajectory $X$ up to time step $k$ therefore belongs to
the space $T_{\left(k\right)}=\uplus_{\left(t,i\right)\in I_{(k)}}\left\{ t\right\} \times\mathbb{R}^{in_{x}}$,
where $\uplus$ stands for disjoint union, which is used to highlight
that the sets are disjoint. Similarly to the set ${\bf x}$ of targets,
we denote a set of trajectories up to time step $k$ as $\mathbf{X}\in\mathcal{F}\left(T_{\left(k\right)}\right)$. 

Given a trajectory $X=\left(t,x^{1:i}\right)$, the set $\tau^{k'}\left(X\right)$,
which can be empty, denotes the corresponding target state at a time
step $k'$. Given a set $\mathbf{X}$ of trajectories, the set $\tau^{k'}\left(\mathbf{X}\right)$
of target states at time $k'$ is $\tau^{k'}\left(\mathbf{X}\right)=\bigcup_{X\in\mathbf{X}}\tau^{k'}\left(X\right)$.

\subsection{Set integral\label{subsec:Set-integral}}

 Given a real-valued function $\pi\left(\cdot\right)$ on the single
trajectory space $T_{\left(k\right)}$, its integral is 
\begin{align}
\int\pi\left(X\right)dX & =\sum_{\left(t,i\right)\in I_{(k)}}\int\pi\left(t,x^{1:i}\right)dx^{1:i}.\label{eq:single_trajectory_integral}
\end{align}
This integral goes through all possible start times, lengths and target
states of the trajectory. Given a real-valued function $\pi\left(\cdot\right)$
on the space $\mathcal{F}\left(T_{\left(k\right)}\right)$ of sets
of trajectories, its set integral is
\begin{align}
\int\pi\left(\mathbf{X}\right)\delta\mathbf{X} & =\sum_{n=0}^{\infty}\frac{1}{n!}\int\pi\left(\left\{ X_{1},...,X_{n}\right\} \right)dX_{1:n}\label{eq:set_integral_trajectory}
\end{align}
where $X_{1:n}=\left(X_{1},...,X_{n}\right)$. Function $\pi\left(\cdot\right)$
is a multitrajectory density if $\pi\left(\cdot\right)\geq0$ and
its set integral is one.

\subsection{Probability hypothesis density\label{subsec:Probability-hypothesis-density}}

The PHD \cite{Mahler_book14} of a multitrajectory density $\pi\left(\cdot\right)$
is 
\begin{align}
D_{\pi}(X) & =\int\pi\left(\left\{ X\right\} \cup\mathbf{X}\right)\delta\mathbf{X}.\label{eq:PHD}
\end{align}
As in the PHD for RFS of targets, integrating the PHD in a region
$A\subseteq T_{\left(k\right)}$ gives us the expected number of trajectories
in this region \cite[Eq. (4.76)]{Mahler_book14}:
\begin{align}
\hat{N}_{A} & =\int_{A}D_{\pi}(X)dX\nonumber \\
 & =\sum_{\left(t,i\right)\in I_{(k)}}\int1_{A}\left(t,x^{1:i}\right)D_{\pi}(t,x^{1:i})dx^{1:i}\label{eq:expected_trajectory_number}
\end{align}
where $1_{A}\left(\cdot\right)$ is the indicator function of a subset
$A$ \cite[App. A.3]{Mahler_book14}. Therefore, the expected number
of trajectories (in total) is given by substituting $A=T_{\left(k\right)}$
into (\ref{eq:expected_trajectory_number}).
\begin{example}
\label{exa:PHD}We consider a multitrajectory density $\nu\left(\cdot\right)$
with
\begin{align}
D_{\nu}\left(1,x^{1}\right) & =\mathcal{N}\left(x^{1};10,1\right)+\mathcal{N}\left(x^{1};1000,1\right)\label{eq:PHD_example1}\\
D_{\nu}\left(1,x^{1:2}\right) & =\mathcal{N}\left(x^{1:2};\left(10,10.1\right),\left[\begin{array}{cc}
1 & 1\\
1 & 2
\end{array}\right]\right),\label{eq:PHD_example2}
\end{align}
where $\mathcal{N}\left(\cdot;m,P\right)$ is a Gaussian density with
mean $m$ and covariance matrix $P$, and zero otherwise. The expected
number of trajectories that start at time one with length 1 is given
by substituting $A=\left\{ 1\right\} \times\mathbb{R}^{n_{x}}$ into
(\ref{eq:expected_trajectory_number}) so
\begin{align*}
\hat{N}_{A} & =\int D_{\nu}\left(1,x^{1}\right)dx^{1}=2.
\end{align*}
The expected number of trajectories is $\hat{N}_{T_{\left(k\right)}}=3$.
$\oblong$
\end{example}

\section{Poisson RFS of trajectories\label{sec:Poisson-trajectory-RFS}}

In this section, we introduce the Poisson RFS on the trajectory space
and some of its properties. 

\subsection{Probability density function\label{subsec:Probability-density-function}}

In the Poisson RFS, the cardinality of the set is Poisson distributed
and its elements are independent and identically distributed (IID).
A Poisson multitrajectory density $\nu\left(\cdot\right)$ has the
form
\begin{align}
\nu\left(\left\{ X_{1},...,X_{n}\right\} \right) & =e^{-\lambda_{\nu}}\lambda_{\nu}^{n}\prod_{j=1}^{n}\breve{\nu}\left(X_{j}\right)\label{eq:Poisson_prior}
\end{align}
where $\breve{\nu}\left(\cdot\right)$ is a single trajectory density,
which implies 
\begin{align*}
\int\breve{\nu}\left(X\right)dX & =1,
\end{align*}
and $\lambda_{\nu}\geq0$. A Poisson multitrajectory density is characterised
by either its PHD $D_{\nu}(X)=\lambda_{\nu}\breve{\nu}\left(X\right)$
or by $\lambda_{\nu}$ and $\breve{\nu}\left(\cdot\right)$ \cite{Mahler_book14}.
As a result, using (\ref{eq:expected_trajectory_number}), the expected
number of trajectories is $\hat{N}_{T_{\left(k\right)}}=\lambda_{\nu}$.
Further, its cardinality distribution is given by \cite{Angel15_prov}
\begin{align}
\rho_{\nu}\left(n\right)=\frac{1}{n!}\int\nu\left(\left\{ X_{1},...,X_{n}\right\} \right)dX_{1:n} & =\frac{1}{n!}e^{-\lambda_{\nu}}\lambda_{\nu}^{n}\label{eq:Poisson_cardinality}
\end{align}
\begin{example}
We consider a Poisson RFS with the PHD of Example \ref{exa:PHD}.
Using (\ref{eq:Poisson_cardinality}), its cardinality distribution
is Poisson with $\lambda_{\nu}=3$ and, therefore, its single trajectory
density is $\breve{\nu}\left(X\right)=D_{\nu}\left(X\right)/3$. $\oblong$
\end{example}
We proceed to explain how to draw samples from $\nu\left(\cdot\right)$.
The probability that a trajectory generated from $\nu\left(\cdot\right)$
starts at time $t$ and has duration $i$ is 
\begin{equation}
P_{\breve{\nu}}\left(t,i\right)=\int\breve{\nu}\left(t,x^{1:i}\right)dx^{1:i}.\label{eq:prob_start_time_duration}
\end{equation}
That is, we integrate over all possible trajectories with start time
$t$ and duration $i$. Given the start time $t$ and duration $i$,
the density of the states is
\begin{align}
\breve{\nu}\left(x^{1:i}|t,i\right) & =\breve{\nu}\left(t,x^{1:i}\right)/P_{\breve{\nu}}\left(t,i\right).\label{eq:conditional_pdf_Poisson}
\end{align}
Therefore, the procedure to draw samples from a Poisson multitrajectory
density $\nu\left(\cdot\right)$ is shown in Algorithm \ref{alg:Sampling_Poisson}.

\begin{algorithm}
\caption{\label{alg:Sampling_Poisson}Sampling from a Poisson multitrajectory
density}

{\fontsize{9}{9}\selectfont

\textbf{Input: }Poisson multitrajectory density $\nu\left(\cdot\right)$. 

\textbf{Output: }Sample $X\thicksim\nu\left(\cdot\right)$.

\begin{algorithmic}     

\State - Set $X=\emptyset$ and sample $n\thicksim\rho_{\nu}\left(\cdot\right)$,
see (\ref{eq:Poisson_cardinality}). 

\For{ $j=1$ to $n$ }  

\State - Sample $\left(t,i\right)\thicksim P_{\breve{\nu}}\left(\cdot\right)$
and $x^{1:i}\thicksim\breve{\nu}\left(\cdot|t,i\right)$, see (\ref{eq:prob_start_time_duration})
and (\ref{eq:conditional_pdf_Poisson}).

\State - Set $X\leftarrow X\cup\left\{ \left(t,x^{1:i}\right)\right\} $. 

\EndFor

\end{algorithmic}

}
\end{algorithm}

\subsection{Marginalisation for Poisson multitrajectory densities\label{subsec:Marginalisation-for-Poisson} }

Given a Poisson multitrajectory density $\nu\left(\cdot\right)$,
the multitarget density $\nu_{\tau}^{k'}\left(\cdot\right)$ of the
targets at a time $k'$, with $1\leq k'\leq k$, is Poisson with PHD
\begin{align}
D_{\nu_{\tau}^{k'}}\left(y\right) & =\sum_{t=1}^{k'}\sum_{j=0}^{k-k'}\int\int D_{\nu}\left(t,x^{1:k'-t},y,z^{1:j}\right)dx^{1:k-t}dz^{1:j}\label{eq:marginalisation_PHD}
\end{align}
where $\left(t,x^{1:k-t},y,z^{1:j}\right)$ denotes a trajectory that
starts at time $t$ with states $\left(x^{1:k-t},y,z^{1:j}\right)$
so it has a duration $k-t+1+j$. This result is obtained from the
basic properties of Poisson processes \cite[Chap. 2]{Kingman_book93}.
\begin{example}
We consider the Poisson multitrajectory density of Example \ref{exa:PHD}.
Using (\ref{eq:marginalisation_PHD}), the set of targets at time
$1$ is Poisson distributed with PHD
\begin{align*}
D_{\nu_{\tau}^{1}}\left(y\right) & =2\mathcal{N}\left(y;10,1\right)+\mathcal{N}\left(y;1000,1\right).
\end{align*}
The expected number of targets at time 1 is 3. $\oblong$
\end{example}

\subsection{KLD minimisation\label{subsec:KLD-minimisation}}

Using FISST \cite{Mahler_book14}, the KLD from $\pi\left(\cdot\right)$
to $\nu\left(\cdot\right)$ is given by 
\begin{align}
\mathrm{D}\left(\pi\left\Vert \nu\right.\right) & =\int\pi\left(\mathbf{X}\right)\log\frac{\pi\left(\mathbf{X}\right)}{\nu\left(\mathbf{X}\right)}\delta\mathbf{X}.\label{eq:KLD_definition}
\end{align}
In Appendix \ref{sec:Appendix_KLD}, we prove the following theorem. 
\begin{thm}
\label{thm:KLD_minimisation_Poisson}Given a multitrajectory density
$\pi\left(\cdot\right)$, the PHD that characterises the Poisson multitrajectory
density $\nu\left(\cdot\right)$ that minimises the KLD $\mathrm{D}\left(\pi\left\Vert \nu\right.\right)$
satisfies $D_{\nu}\left(\cdot\right)=D_{\pi}\left(\cdot\right)$.
\end{thm}

\section{Trajectory PHD filter\label{sec:Trajectory-PHD-filter}}

In this section, we derive the TPHD filter. In Section \ref{subsec:Bayesian-recursion},
we present the Bayesian filtering recursion for sets of trajectories.
The prediction and update steps of the TPHD filter are given in Sections
\ref{subsec:Prediction} and \ref{subsec:Update}, respectively. 

\subsection{Bayesian filtering recursion\label{subsec:Bayesian-recursion}}

The objective is to calculate the multitrajectory filtering density
$\pi^{k}\left(\cdot\right)$ at time $k$, which is the multitrajectory
density of the set of trajectories up to time step $k$ conditioned
on the measurements up to time step $k$. We assume that the set of
trajectories at time $k$ evolves with a transition density $f^{k}\left(\cdot\left|\cdot\right.\right)$.
In addition, given the targets at time $k$, the set $\mathbf{z}^{k}$
of measurements at time $k$ has a density $\ell^{k}\left(\cdot|\tau^{k}\left(\mathbf{X}\right)\right)$.
We can calculate $\pi^{k}\left(\cdot\right)$ via the prediction and
update steps \cite{Angel15_prov}:
\begin{align}
\omega^{k}\left(\mathbf{X}\right) & =\int f^{k}\left(\mathbf{X}\left|\mathbf{Y}\right.\right)\pi^{k-1}\left(\mathbf{Y}\right)\delta\mathbf{Y}\label{eq:prediction_trajectories}\\
\pi^{k}\left(\mathbf{X}\right) & =\frac{\ell^{k}\left(\mathbf{z}^{k}|\tau^{k}\left(\mathbf{X}\right)\right)\omega^{k}\left(\mathbf{X}\right)}{\ell^{k}\left(\mathbf{z}^{k}\right)}\label{eq:update_trajectories}
\end{align}
where $\omega^{k}\left(\cdot\right)$ is the predicted multitrajectory
density at time $k$, which denotes the density of the set of trajectories
up to time step $k$ given the measurements up to time step $k-1$.
The density of the measurements is 
\begin{align*}
\ell^{k}\left(\mathbf{z}^{k}\right) & =\int\ell^{k}\left(\mathbf{z}^{k}|\tau^{k}\left(\mathbf{X}\right)\right)\omega^{k}\left(\mathbf{X}\right)\delta\mathbf{X}.
\end{align*}

\subsection{Prediction\label{subsec:Prediction}}

We make the following assumptions in the prediction step: 
\begin{itemize}
\item P1 Given the current multitarget state ${\bf x}$, each target $x\in{\bf x}$
survives with probability $p_{S}\left(x\right)$ and moves to a new
state with a transition density $g\left(\cdot\left|x\right.\right)$,
or dies with probability $1-p_{S}\left(x\right)$.
\item P2 The multitarget state at the next time step is the union of the
surviving targets and new targets, which are born independently with
a Poisson multitarget density $\beta_{\tau}\left(\cdot\right)$.
\item P3 The multitrajectory density $\pi^{k-1}\left(\cdot\right)$ is Poisson.
\end{itemize}
Let $\mathbb{N}_{k}=\left\{ 1,...,k\right\} $. Then, the relation
between predicted PHD at time $k$ and the PHD of the posterior at
time $k-1$ is given by the following theorem, which is proved in
Appendix \ref{sec:Append_TPHD_prediction}. 
\begin{thm}[Prediction]
\label{thm:PHD_prediction}Under Assumptions P1-P3, the predicted
PHD $D_{\omega^{k}}\left(\cdot\right)$ at time $k$ is
\begin{align*}
D_{\omega^{k}}\left(X\right) & =D_{\xi^{k}}\left(X\right)+D_{\beta^{k}}\left(X\right)
\end{align*}
where
\begin{align*}
D_{\beta^{k}}\left(t,x^{1:i}\right) & =D_{\beta_{\tau}}\left(x^{1}\right)1_{\left\{ k\right\} }\left(t\right)1_{\left\{ 1\right\} }\left(i\right)
\end{align*}
\begin{align*}
 & D_{\xi^{k}}\left(t,x^{1:i}\right)\\
 & \,=D_{\pi^{k-1}}\left(t,x^{1:i}\right)1_{\mathbb{N}_{k-2}}\left(t+i-1\right)+\left(1-p_{S}\left(x^{i}\right)\right)\\
 & \quad\times D_{\pi^{k-1}}\left(t,x^{1:i}\right)1_{\left\{ k-1\right\} }\left(t+i-1\right)+p_{S}\left(x^{i-1}\right)\\
 & \quad\times g\left(x^{i}\left|x^{i-1}\right.\right)D_{\pi_{k-1}}\left(t,x^{1:i-1}\right)1_{\left\{ k\right\} }\left(t+i-1\right).
\end{align*}
 
\end{thm}
The predicted PHD is the sum of the PHD $D_{\beta^{k}}\left(\cdot\right)$
of the trajectories born at time step $k$ and the PHD $D_{\xi^{k}}\left(\cdot\right)$
of the trajectories present at previous time steps. The end time of
trajectory $\left(t,x^{1:i}\right)$ is $t+i-1$. Therefore, the three
terms of $D_{\xi^{k}}\left(\cdot\right)$ have clear interpretations
in terms of trajectory end times. The prediction step does not change
the PHD for the trajectories that ended before time step $k-1$. The
PHD of the trajectories that end at time step $k-1$ is multiplied
by $1-p_{S}\left(x^{i}\right)$, which represents the probability
of not surviving. For the surviving trajectories, we multiply the
PHD by the transition density and the survival probability. 

\subsection{Update\label{subsec:Update}}

We make the following assumptions in the update step \cite{Angel15_d}:
\begin{itemize}
\item U1 For a given multi-target state $\mathbf{x}$ at time $k$, each
target state $x\in\mathbf{x}$ is either detected with probability
$p_{D}\left(x\right)$ and generates one measurement with density
$l\left(\cdot|x\right)$, or missed with probability $1-p_{D}\left(x\right)$. 
\item U2 The measurement $\mathbf{z}^{k}$ is the union of the target-generated
measurements and Poisson clutter with density $c\left(\cdot\right)$. 
\item U3 The multitrajectory density $\omega{}^{k}\left(\cdot\right)$ is
Poisson.
\end{itemize}
Let $\Xi_{n,n_{z}}$ denote the set that contains all the vectors
$\sigma=\left(\sigma_{1},...,\sigma_{n}\right)$ that indicate associations
of $n_{z}$ measurements to $n$ targets, which can be either detected
or undetected. If $\sigma\in\Xi_{n,n_{z}}$, $\sigma_{i}=j\in\left\{ 1,...,n_{z}\right\} $
indicates measurement $j$ is associated with target $i$ and $\sigma_{i}=0$
indicates that target $i$ has not been detected. Under Assumptions
U1 and U2, which define the standard measurement model, the density
of the measurement given the state is \cite[Eq. (7.21)]{Mahler_book14}
\begin{align}
 & \ell^{k}\left(\left\{ z_{1},...,z_{n_{z}}\right\} \left|\left\{ x_{1},...,x_{n}\right\} \right.\right)\nonumber \\
 & \quad=e^{-\lambda_{c}}\left[\prod_{i=1}^{n_{z}}\lambda_{c}\breve{c}\left(z_{i}\right)\right]\left[\prod_{i=1}^{n}\left(1-p_{D}\left(x_{i}\right)\right)\right]\nonumber \\
 & \qquad\times\sum_{\sigma\in\Xi_{n,n_{z}}}\prod_{i:\sigma_{i}>0}\frac{p_{D}\left(x_{i}\right)l\left(z_{\sigma_{i}}|x_{i}\right)}{\left(1-p_{D}\left(x_{i}\right)\right)\lambda_{c}\breve{c}\left(z_{\sigma_{i}}\right)}.\label{eq:PDF_measurement_full_PHD}
\end{align}
where $\lambda_{c}$ and $\breve{c}\left(\cdot\right)$ characterise
$c\left(\cdot\right)$, see (\ref{eq:Poisson_prior}). 

Let $L_{\mathbf{z}^{k}}\left(\cdot\right)$ denote the PHD filter
pseudolikelihood function, which is given by \cite[Sec. 8.4.3]{Mahler_book14}
\begin{align*}
L_{\mathbf{z}^{k}}\left(x\right) & =1-p_{D}\left(x\right)+p_{D}\left(x\right)\\
 & \quad\times\sum_{z\in\mathbf{z}^{k}}\frac{l\left(z|x\right)}{\lambda_{c}\breve{c}\left(z\right)+\int p_{D}\left(y\right)l\left(z|y\right)D_{\omega_{\tau}^{k}}\left(y\right)dy}
\end{align*}
with $D_{\omega_{\tau}^{k}}\left(\cdot\right)$ representing the PHD
of the targets at time $k$ of density $\omega^{k}\left(\cdot\right)$,
see (\ref{eq:marginalisation_PHD}):
\begin{align*}
D_{\omega_{\tau}^{k}}\left(y\right) & =\sum_{t=1}^{k}\int D_{\omega^{k}}\left(t,x^{1:k-t},y\right)dx^{1:k-t}.
\end{align*}
Then, we prove in Appendix \ref{sec:Append_TPHD_update} the TPHD
filter update step: 
\begin{thm}[Update]
\label{thm:PHD_update}Under Assumptions U1-U3, the updated PHD $D_{\pi^{k}}\left(\cdot\right)$
at time $k$ is
\begin{align*}
D_{\pi^{k}}\left(t,x^{1:i}\right) & =D_{\omega^{k}}\left(t,x^{1:i}\right)\left[1_{\mathbb{N}_{k-1}}\left(t+i-1\right)\right.\\
 & \quad\left.+1_{\left\{ k\right\} }\left(t+i-1\right)L_{\mathbf{z}^{k}}\left(x^{i}\right)\right]
\end{align*}
if $t+i-1\leq k$ or zero otherwise.
\end{thm}
As in the prediction step, the update step does not change the PHD
for the trajectories that have died before time step $k$. It should
be noted that Bayes update (\ref{eq:update_trajectories}) uses a
likelihood (\ref{eq:PDF_measurement_full_PHD}) which involves a summation
over all target to measurements associations in the multitarget space.
In contrast, the TPHD filter update is similar to the PHD filter update
in the sense that it uses a pseudolikelihood function $L_{\mathbf{z}^{k}}\left(\cdot\right)$,
which is defined on the single target space and only involves associations
between a single target and the measurements. 

It can be checked that if we perform marginalisation at time step
$k$, see (\ref{eq:marginalisation_PHD}), and apply the (target)
PHD prediction and update, we obtain the same result as if we apply
the (trajectory) PHD prediction and update and then apply marginalisation.
Consequently, the information regarding the set of targets at the
current time step is the same for the PHD and TPHD filters. For example,
the estimated cardinality of alive trajectories/targets is the same
for both filters.

\section{Gaussian mixture TPHD filter\label{sec:Gaussian-mixture-TPHD-filter}}

In this section, we propose a Gaussian mixture implementation of the
TPHD filter. The prediction and update steps are provided in Section
\ref{subsec:Prediction-and-update}. We motivate why the GMTPHD should
only be used to track alive trajectories in Section \ref{subsec:GMTPHD_Discussion}.
The $L$-scan GMTPHD, which is a computationally efficient implementation,
is described in Section \ref{subsec:L-scan-GMTPHD}. An estimation
procedure for the GMTPHD filter is given in Section \ref{subsec:Estimation}. 

\subsection{Prediction and update\label{subsec:Prediction-and-update}}

The recursion of the GMTPHD filter is quite similar to the GMPHD filter
\cite{Vo06}. We use the notation 
\begin{align}
\mathcal{N}\left(t,x^{1:i};t^{k},m^{k},P^{k}\right) & =\mathcal{N}\left(x^{1:i};m^{k},P^{k}\right)\quad\mathrm{if}\:i=i^{k},t=t^{k}\label{eq:Trajectory_Gaussian}
\end{align}
or zero otherwise, where $i^{k}=\mathrm{dim}\left(m^{k}\right)/n_{x}$.
Equation (\ref{eq:Trajectory_Gaussian}) represents a single trajectory
Gaussian density with start time $t^{k}$, duration $i^{k}$, mean
$m^{k}\in\mathbb{R}^{i^{k}n_{x}}$ and covariance matrix $P^{k}\in\mathbb{R}^{i^{k}n_{x}\times i^{k}n_{x}}$
evaluated at $\left(t,x^{1:i}\right)$. We use $\otimes$ to indicate
Kronecker product and $0_{m,n}$ is the $m\times n$ zero matrix.

We make the additional assumptions
\begin{itemize}
\item A1 The probabilities $p_{S}$ and $p_{D}$ are constants. 
\item A2 $g\left(x^{i}\left|x^{i-1}\right.\right)=\mathcal{N}\left(x^{i};Fx^{i-1},Q\right)$. 
\item A3 $l\left(z|x\right)=\mathcal{N}\left(z;Hx,R\right)$.
\item A4 The PHD of the birth density $\beta^{k}\left(\cdot\right)$ is
\begin{align}
D_{\beta^{k}}\left(X\right) & =\sum_{j=1}^{J_{\beta}^{k}}w_{\beta,j}^{k}\mathcal{N}\left(X;k,m_{\beta,j}^{k},P_{\beta,j}^{k}\right)\label{eq:GMPHD-birth}
\end{align}
where $J_{\beta}^{k}\in\mathbb{N}$ is the number of components, $m_{\beta,j}^{k}\in\mathbb{R}^{n_{x}}$
and $P_{\beta,j}^{k}\in\mathbb{R}^{n_{x}\times n_{x}}$. 
\end{itemize}
It should be noted that the models provided by A1-A4 could be time
varying but omit time for notational convenience. Under Assumptions
A1-A4, P1-P3 and U1-U3, we can calculate the TPHD filter in closed
form giving rise to the GMTPHD filter, whose prediction and update
steps are provided in the following. 
\begin{prop}[Prediction]
\label{prop:GMTPHD_prediction}We denote the PHD of $\pi^{k}\left(\cdot\right)$
by
\begin{align*}
D_{\pi^{k}}\left(X\right) & =D_{\pi_{\star}^{k}}\left(X\right)+D_{\pi_{\circ}^{k}}\left(X\right)
\end{align*}
where
\begin{align*}
D_{\pi_{\star}^{k}}\left(X\right) & =\sum_{j=1}^{J^{k}}w_{j}^{k}\mathcal{N}\left(X;t_{j}^{k},m_{j}^{k},P_{j}^{k}\right)\\
D_{\pi_{\circ}^{k}}\left(X\right) & =\sum_{j=1}^{J_{\circ}^{k}}w_{\circ,j}^{k}\mathcal{N}\left(X;t_{\circ,j}^{k},m_{\circ,j}^{k},P_{\circ,j}^{k}\right)
\end{align*}
represent the PHD of alive and dead trajectories, i.e., $t_{j}^{k}+i_{j}^{k}-1=k$
and $t_{\circ,j}^{k}+i_{\circ,j}^{k}-1<k$ with $i_{j}^{k}=\mathrm{dim}\left(m_{j}^{k}\right)/n_{x}$
and $i_{\circ,j}^{k}=\mathrm{dim}\left(m_{\circ,j}^{k}\right)/n_{x}$.
Then, the PHD of $\omega^{k+1}\left(\cdot\right)$ is
\begin{align}
D_{\omega^{k+1}}\left(X\right) & =\left(1-p_{S}\right)D_{\pi_{\star}^{k}}\left(X\right)+D_{\pi_{\circ}^{k}}\left(X\right)+D_{\beta^{k+1}}\left(X\right)\nonumber \\
 & \quad+p_{S}\sum_{j=1}^{J^{k}}w_{j}^{k}\mathcal{N}\left(X;t_{j}^{k},m_{\omega,j}^{k+1},P_{\omega,j}^{k+1}\right)\label{eq:Prop7_equation}
\end{align}
where
\begin{align*}
m_{\omega,j}^{k+1} & =\left[\left(m_{j}^{k}\right)^{T},\left(\dot{F}_{j}m_{j}^{k}\right)^{T}\right]^{T}\\
P_{\omega,j}^{k+1} & =\left[\begin{array}{cc}
P_{j}^{k} & P_{j}^{k}\dot{F}_{j}^{T}\\
\dot{F}_{j}P_{j}^{k} & \dot{F}_{j}P_{j}^{k}\dot{F}_{j}^{T}+Q
\end{array}\right]\\
\dot{F}_{j} & =\left[0_{1,i_{j}^{k}-1},1\right]\otimes F.
\end{align*}
\end{prop}
Proposition \ref{prop:GMTPHD_prediction} can be proved using Theorem
\ref{thm:PHD_prediction}. The GMTPHD filter prediction is similar
to the GMPHD filter prediction with the main differences that previous
states are not integrated out, as in \cite{Koch11}, and there is
information about dead trajectories.
\begin{prop}[Update]
\label{prop:GMTPHD_update}We denote the PHD of $\omega^{k}\left(\cdot\right)$
by
\begin{align*}
D_{\omega^{k}}\left(X\right) & =D_{\omega_{\star}^{k}}\left(X\right)+D_{\omega_{\circ}^{k}}\left(X\right)
\end{align*}
where
\begin{align*}
D_{\omega_{\star}^{k}}\left(X\right) & =\sum_{j=1}^{J_{\omega}^{k}}w_{\omega,j}^{k}\mathcal{N}\left(X;t_{\omega,j}^{k},m_{\omega,j}^{k},P_{\omega,j}^{k}\right)\\
D_{\omega_{\circ}^{k}}\left(X\right) & =\sum_{j=1}^{J_{\circ}^{k}}w_{\circ,j}^{k}\mathcal{N}\left(X;t_{\circ,j}^{k},m_{\circ,j}^{k},P_{\circ,j}^{k}\right)
\end{align*}
represent the PHD of alive and dead trajectories. Then, the PHD of
$\pi^{k}\left(\cdot\right)$ is
\begin{align}
D_{\pi^{k}}\left(X\right) & =D_{\omega_{\circ}^{k}}\left(X\right)+\left(1-p_{D}\right)D_{\omega_{\star}^{k}}\left(X\right)\nonumber \\
 & \,+\sum_{z\in\mathbf{z}^{k}}\sum_{j=1}^{J^{k}}w_{j}\left(z\right)\mathcal{N}\left(X;t_{\omega,j}^{k},m_{j}^{k}\left(z\right),P_{j}^{k}\right)\label{eq:Prop8_equation}
\end{align}
where
\begin{align*}
w_{j}\left(z\right) & =\frac{p_{D}w_{\omega,j}^{k}\mathcal{N}\left(z;\overline{z}_{j},S_{j}\right)}{\lambda_{c}\breve{c}\left(z\right)+p_{D}\sum_{l=1}^{J_{\omega}^{k}}w_{\omega,l}^{k}\mathcal{N}\left(z;\overline{z}_{l},S_{l}\right)}\\
\overline{z}_{j} & =\dot{H}_{j}m_{\omega,j}^{k},\quad S_{j}=\dot{H}_{j}P_{\omega,j}^{k}\dot{H}_{j}^{T}+R\\
\dot{H}_{j} & =\left[0_{1,i_{\omega,j}^{k}-1},1\right]\otimes H\\
m_{j}^{k}\left(z\right) & =m_{\omega,j}^{k}+P_{\omega,j}^{k}\dot{H}^{T}S_{j}^{-1}\left(z-\overline{z}_{j}\right)\\
P_{j}^{k} & =P_{\omega,j}^{k}-P_{\omega,j}^{k}\dot{H}^{T}S_{j}^{-1}\dot{H}P_{\omega,j}^{k}.
\end{align*}
where $i_{\omega,j}^{k}=\mathrm{dim}\left(m_{\omega,j}^{k}\right)/n_{x}$
\end{prop}
Proposition \ref{prop:GMTPHD_update} can be proved using Theorem
\ref{thm:PHD_update}. As $D_{\omega_{\star}^{k}}\left(\cdot\right)$
and $D_{\omega_{\circ}^{k}}\left(\cdot\right)$ represent the alive
and dead trajectories, respectively, it is met that $t_{\omega,j}^{k}+i_{\omega,j}^{k}-1=k$
and $t_{\circ,j}^{k}+i_{\circ,j}^{k}-1<k$. Also, the GMTPHD filter
update is similar to the GMPHD filter update. The main differences
is that we keep the PHD that represents dead trajectories and we update
the whole trajectories. The updated weights of the alive components
are the same as in the GMPHD filter because the likelihood only depends
on the the current set of targets.

\subsection{Tracking of only alive trajectories\label{subsec:GMTPHD_Discussion}}

In this section, we motivate why practical GMTPHD implementations
should not attempt to track the dead trajectories. As in the PHD filter,
the Poisson approximation for the multitarget density of the current
set of targets, is a strong approximation but yields acceptable results
in many situations \cite{Mahler_book14}. The Poisson approximation
for the multitrajectory density is even stronger as we proceed to
explain. First, the number of total trajectories is, in most cases,
greater than the number of current targets and, therefore, the Poisson
approximation for the number of trajectories is usually worse (the
variance of a Poisson distribution is equal to its mean). Second,
in practice, we argue that the Poisson approximation is only useful
to obtain information about the present trajectories at the current
time step. The reason is that, in the prediction step, the weight
of the components of trajectories that die at the current time step
is multiplied by $\left(1-p_{S}\right)$, see the first term in (\ref{eq:Prop7_equation}).
Then, the weights of the components of dead trajectories are never
modified at future time steps, see $D_{\pi_{\circ}^{k}}\left(X\right)$
and $D_{\omega_{\circ}^{k}}\left(X\right)$ in Propositions \ref{prop:GMTPHD_prediction}
and \ref{prop:GMTPHD_update}. The probability $p_{S}$ of survival
is usually close to one so these components have very low weights.
As a result, all components that represent dead trajectories have
very low weight even if they were very likely in the past. 

The conclusion is that the Poisson approximation to the full multitrajectory
filtering density is not an accurate representation of the knowledge
over all trajectories that have existed up to the current time. 
Nevertheless, the TPHD filter is useful to approximate the posterior
of the alive trajectories. In practice, this implies setting $D_{\pi_{\circ}^{k}}\left(X\right)=0$
and removing the term $\left(1-p_{S}\right)D_{\pi_{\star}^{k}}\left(X\right)$
in (\ref{eq:Prop7_equation}) and setting $D_{\omega_{\circ}^{k}}\left(X\right)=0$
in (\ref{eq:Prop8_equation}).

\subsection{$L$-scan GMTPHD\label{subsec:L-scan-GMTPHD}}

In this section, we propose a computationally efficient implementation
of the GMTPHD filter: the $L$-scan GMTPHD filter. The GMTPHD filter
has an increasing number of components as time progresses so we need
to bound the number of components in practice. The simplest technique
is to prune the components whose weight is below a threshold $\Gamma_{p}$
and set a maximum number $J_{max}$ of components \cite{Vo06}. In
addition, if two components have a very similar current state, based
on a Mahalanobis distance criterion, future measurements will affect
both component weights and future states in a similar way. Therefore,
we can remove components that are close to another component with
higher weight. We account for this decrease in the number of components
by increasing the weight of the component that has not been removed
by the weights of the removed components. We refer to as this technique
as absorption. The steps of the pruning and absorption algorithms
for the GMTPHD are given in Algorithm \ref{alg:Pruning-and-absorption},
where we use the notation $\Phi_{j}^{k}=\left(w_{j}^{k},t_{j}^{k},m_{j}^{k},P_{j}^{k}\right)$
. 

\begin{algorithm}
\caption{\label{alg:Pruning-and-absorption}Pruning and absorption for the
GMTPHD filter}

{\fontsize{9}{9}\selectfont 

\textbf{Input: }Posterior parameters $\left\{ \Phi_{j}^{k}\right\} _{j=1}^{J^{k}}$,
pruning threshold $\Gamma_{p}$, absorption threshold $\Gamma_{a}$,
maximum number of terms $J_{max}$. 

\textbf{Output:} Pruned posterior parameters $\left\{ \Phi_{o,j}^{k}\right\} _{j=1}^{\hat{J}^{k}}$

\begin{algorithmic}     

\State - Set $l=0$ and $I=\left\{ j\in\left\{ 1,...,J^{k}\right\} :w_{j}^{k}>\Gamma_{p}\right\} $.

\While{$I\neq\emptyset$}

\State - Set $l\leftarrow l+1$.

\State - $j=\underset{i\in I}{\arg\max}\:w_{i}^{k}$.

\State - $L=\left\{ i\in I:\left(\hat{m}_{i}^{k}-\hat{m}_{j}^{k}\right)^{T}\left(\hat{P}_{j}^{k}\right)^{-1}\left(\hat{m}_{i}^{k}-\hat{m}_{j}^{k}\right)\leq\Gamma_{a}\right\} $
with $\hat{m}_{j}^{k}\in\mathbb{R}^{n_{x}}$ and $\hat{P}_{j}^{k}\in\mathbb{R}^{n_{x}\times n_{x}}$
denoting the mean and covariance matrix of the state at the current
time step.

\State - $\Phi_{o,l}^{k}=\Phi_{j}^{k}$ with weight $w_{o,l}^{k}=\sum_{i\in L}w_{i}^{k}$.

\State - $I\leftarrow I\setminus L$.

\EndWhile

\State - If $l>J_{max}$, only keep the $J_{max}$ components with
highest weight

\end{algorithmic}

}
\end{algorithm}

In addition, as time progresses, the lengths of the trajectories increase
so, eventually, the direct implementation of the GMTPHD is not computationally
feasible. Fortunately, in practice, measurements at the current time
step only have a significant impact on the trajectory state estimates
for recent time steps. Based on this insight combined with the ADF
framework and KLD minimisation, we propose a computationally efficient,
single trajectory $L$-scan filter in Appendix \ref{sec:Append_efficient_implementation}.
The density that this filter propagates is composed by the joint density
of the states of the last $L$ time steps and independent densities
for the previous states. We apply this filter to each mixture component
of the GMTPHD posterior and the resulting algorithm is referred to
as $L$-scan GMTPHD. 

The $L$-scan GMTPHD is implemented as the GMTPHD with a minor modification
in the prediction step, where we discard the correlations of states
that happened at least $L$ time steps before the current time step.
Given a predicted PHD $D_{\omega^{k}}\left(\cdot\right)$, see Proposition
\ref{prop:GMTPHD_prediction}, its $L$-scan version is
\begin{align}
D_{\omega^{k}}^{(L)}\left(X\right) & =\sum_{j=1}^{J_{\omega}^{k}}w_{\omega,j}^{k}\mathcal{N}\left(X;t_{\omega,j}^{k},m_{\omega,j}^{k},P_{\omega,j}^{k(L)}\right)\label{eq:Lscan_prediction}
\end{align}
where $P_{\omega,j}^{k(L)}=\mathrm{diag}\left(\tilde{P}_{j}^{t_{\omega,j}^{k}},\tilde{P}_{j}^{t_{\omega,j}^{k}+1},...,\tilde{P}_{j}^{k-L},\tilde{P}_{j}^{k-L+1:k}\right)$.
Matrix $\tilde{P}_{j}^{k-L+1:k}\in\mathbb{R}^{L\cdot n_{x}\times L\cdot n_{x}}$
represents the joint covariance of the $L$ last time instants, obtained
from $P_{\omega,j}^{k}$, and $\tilde{P}_{j}^{k}\in\mathbb{R}^{n_{x}\times n_{x}}$
represents the covariance matrix of the target state at time $k$,
obtained from $P_{\omega,j}^{k}$. Therefore, we have independent
Gaussian densities to represent the states outside the $L$-scan window
and a joint Gaussian density for the states in the $L$-scan window.
The steps of the $L$-scan GMTPHD filter are summarised in Algorithm
\ref{alg:Lscan_GMTPHD_steps}. 

It should be noted that the estimated number of alive trajectories
and the target states at the current time are not affected by  $L$.
This implies that the estimated number of alive trajectories is equal
to the number of targets of the GMPHD filter and the estimated targets
at the current time using both the GMPHD or GMTPHD are alike. 

\begin{algorithm}
\caption{\label{alg:Lscan_GMTPHD_steps}$L$-scan GMTPHD filter steps}

{\fontsize{9}{9}\selectfont

\begin{algorithmic}     

\State - Initialisation: $D_{\pi^{0}}\left(\cdot\right)=0$: $J^{0}=0$.

\For{ $k=1$ to \textit{final time step} }

\State - Prediction using Proposition \ref{prop:GMTPHD_prediction}
with this modification:

\State $\quad$- After calculating $P_{\omega,j}^{k}$, represent
it in the form of $P_{\omega,j}^{k(L)}$, see (\ref{eq:Lscan_prediction}),
by discarding correlations outside the $L$-scan window. 

\State - Update using Proposition \ref{prop:GMTPHD_update}. 

\State - Estimation of the alive trajectories, see Section \ref{subsec:Estimation}.

\State - Pruning/absorption using Algorithm \ref{alg:Pruning-and-absorption}.

\EndFor

\end{algorithmic}

}
\end{algorithm}

\subsection{Estimation\label{subsec:Estimation}}

We adapt the estimator for the GMPHD filter described in \cite[Sec. 9.5.4.4]{Mahler_book14}
for sets of trajectories. First, the number of trajectories is estimated
as
\begin{align}
\hat{N}^{k} & =\mathrm{round}\left(\sum_{j=1}^{J^{k}}w_{j}^{k}\right).\label{eq:estimated_trajectories}
\end{align}
Then, the estimated set of trajectories corresponds to $\left\{ \left(t_{l_{1}}^{k},m_{l_{1}}^{k}\right),...,\left(t_{l_{\hat{N}^{k}}}^{k},m_{l_{\hat{N}^{k}}}^{k}\right)\right\} $
where $\left\{ l_{1},...,l_{\hat{N}^{k}}\right\} $ are the indices
of the components with highest weights. 

There are several drawbacks with this sub-optimal estimator. First,
$J^{k}$ cannot be smaller than $\hat{N}^{k}$. Also, this estimator
does not work well if there is a component with weight higher than
two because there are at least two targets in that region but only
one is reported. Nevertheless, this estimator is commonly used in
the GMPHD filter and has a low computational complexity, so we suggest
its use for the GMTPHD filter as well.

\section{Simulations\label{sec:Simulations}}

 We proceed to assess the performance of the $L$-scan TPHD filter
by simulations. We consider a target state $x=\left[p_{x},\dot{p}_{x},p_{y},\dot{p}_{y}\right]^{T}$,
which contains position and velocity. All the units of the quantities
in this section are given in the international system. The parameters
of the single-target dynamic process are 
\begin{align*}
F=I_{2}\otimes\left(\begin{array}{cc}
1 & \tau\\
0 & 1
\end{array}\right),\quad Q=qI_{2}\otimes\left(\begin{array}{cc}
\tau^{3}/3 & \tau^{2}/2\\
\tau^{2}/2 & \tau
\end{array}\right)
\end{align*}
where $\tau=0.5$ is the sampling time and $q=3.24$ is a parameter.
We also set $p_{S}=0.99$. The parameters of the measurement model
are 
\begin{align*}
H=\left(\begin{array}{cccc}
1 & 0 & 0 & 0\\
0 & 0 & 1 & 0
\end{array}\right),\quad R=\sigma^{2}I_{2},
\end{align*}
where $\sigma^{2}=16$, and $p_{D}=0.9$. The clutter intensity is
$D_{c}\left(z\right)=\lambda_{c}\cdot u_{A}\left(z\right)$ where
$u_{A}\left(z\right)$ is a uniform density in region $A=\left[0,2000\right]\times\left[0,2000\right]$
and $\lambda_{c}=50$ is the average number of clutter measurements
per scan. The birth process parameters are $J_{\beta}^{k}=3$, $w_{\beta,j}^{k}=0.1$,
$P_{\beta,j}^{k}=100I_{4}$ for $j\in\left\{ 1,2,3\right\} $ and
$m_{\beta,1}^{k}=\left[85,0,140,0\right]^{T}$, $m_{\beta,2}^{k}=\left[-5,0,220,0\right]^{T}$
and $m_{\beta,3}^{k}=\left[7,0,50,0\right]^{T}$. 

\begin{figure}
\begin{centering}
\includegraphics[scale=0.31]{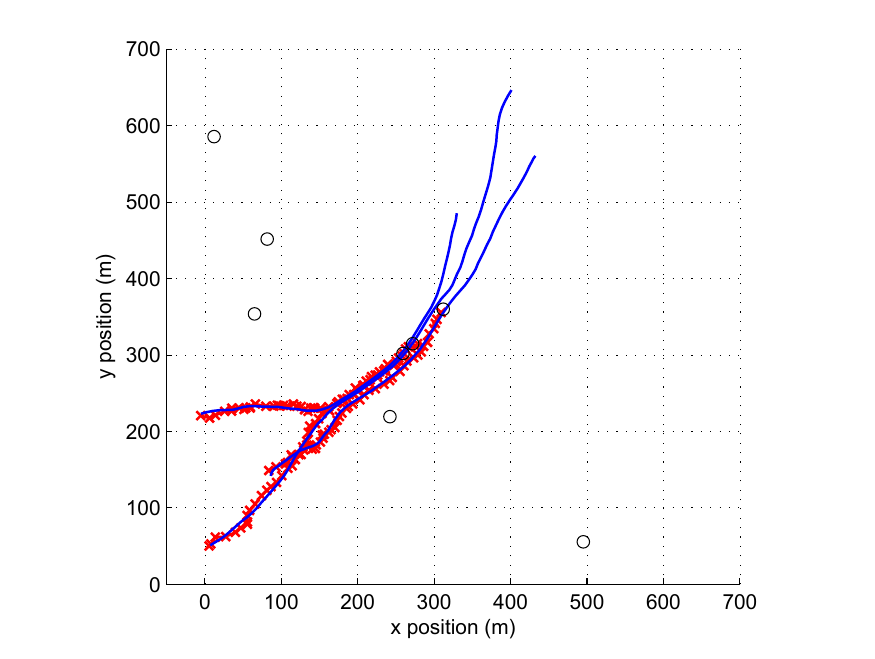}\includegraphics[scale=0.31]{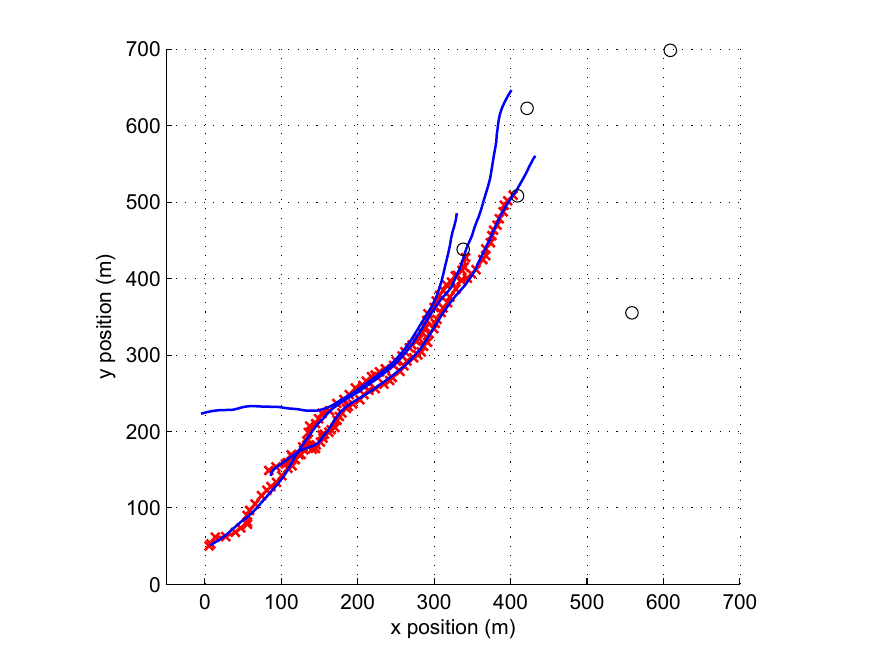}
\par\end{centering}
\caption{\label{fig:Exemplar-outputs}Exemplar outputs at time steps 50 (left)
and 70 (right). The blue lines represent the true trajectories, which
start at time steps $\left(1,5,10\right)$ and finish at $\left(80,70,95\right)$.
The red lines with crosses represent the estimated alive trajectories.
Black circles represent the current measurements. The TPHD filter
is able to estimate the alive trajectories.}
\end{figure}

We have implemented the $L$-scan TPHD filter with $L\in\left\{ 1,2,5,10\right\} $
in a scenario with 100 time steps. We use a pruning threshold $\Gamma_{p}=10^{-4}$,
absorption threshold $\Gamma_{a}=4$ and limit the number of components
to 30. Two exemplar outputs of the $10$-scan TPHD filter and the
considered ground truth are shown in Figure \ref{fig:Exemplar-outputs}.
At each time step, the TPHD provides an estimate of the set of present
trajectories at the current time. The start and end times of an estimated
trajectory do not depend on the choice of $L$ so the output for any
other $L$ looks alike but with a different error. 

\begin{figure}
\begin{centering}
\includegraphics[scale=0.5]{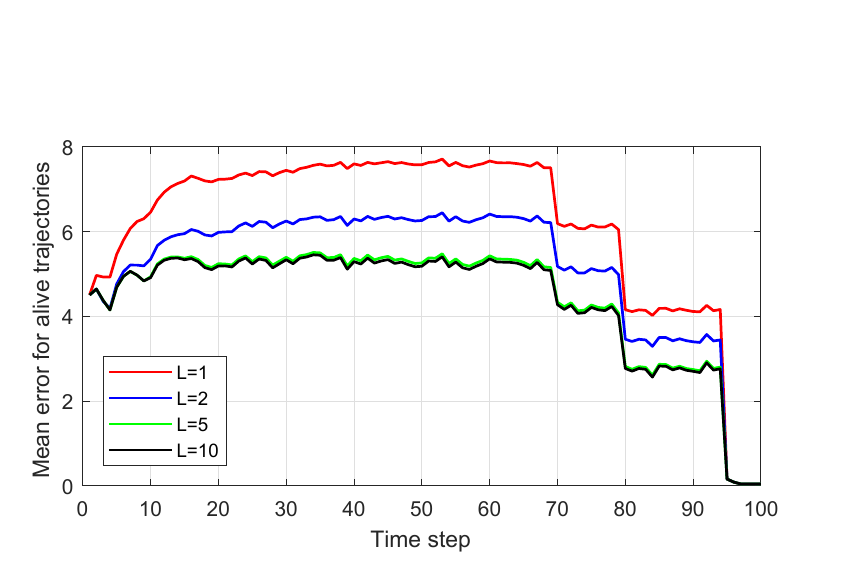}
\par\end{centering}
\caption{\label{fig:Averaged-OSPA-error}Error calculated with the sets of
trajectories metric for the $L$-scan TPHD filter. Performance improves
as we increase $L$. }
\end{figure}

In the following, we evaluate the performance of the $L$-scan TPHD
filters by Monte Carlo simulation with 500 runs. At each time step
$k$, we measure the distance between the set $\mathbf{X}_{a}^{k}$
of alive trajectories and its estimate $\mathbf{\hat{X}}_{a}^{k}$
using the metric $d\left(\cdot,\cdot\right)$ for sets of trajectories
based on linear programming in \cite{Rahmathullah16_prov2}, with
parameters $p=2$, $c=10$ and $\gamma=0.1$. We only use the position
elements and normalise the metric by $\sqrt{k}$. The resulting mean
errors for the $L$-scan TPHD filter are plotted in Figure \ref{fig:Averaged-OSPA-error}.
At the beginning, the filters have the same error but soon the differences
start to appear. As expected, the error decreases as we increase $L$
in the filter because we are considering a longer time window to update
the trajectories. In addition, the running times of a non-optimised
Matlab implementation on a Intel Core i7 laptop are basically the
same for $L\in\left\{ 1,2,5,10\right\} $: 7.7 seconds. In our implementation,
the computational burden associated to operations resulting of using
an $L$ window of sizes between 1 and 10 is negligible compared to
the computational burden of the rest of the filter. If we continue
increasing $L$, the running time increases considerable, for example,
14.8 s for $L=20$ and 28.0 s for $L=30$. 

We also show the error averaged over all time steps changing several
parameters of the simulation in Table \ref{tab:Cost-function-different_parameters}.
Logically, with lower measurement noise or clutter rate or higher
probability of detection, performance increases. 

\begin{table}
\caption{\label{tab:Cost-function-different_parameters}Error in alive trajectories
averaged over all time steps}
\begin{centering}
\begin{tabular}{c|cccc}
\hline 
Changed parameter &
$L$=1  &
$L$=2 &
$L$=5 &
$L$=10\tabularnewline
\hline 
No change &
6.20 &
5.18 &
4.46 &
4.41\tabularnewline
$\sigma^{2}=25$ &
7.12 &
6.02 &
5.08 &
5.02\tabularnewline
$\sigma^{2}=9$ &
5.20 &
4.32 &
3.82 &
3.79\tabularnewline
$\lambda_{c}=70$ &
6.25 &
5.24 &
4.52 &
4.47\tabularnewline
$\lambda_{c}=90$ &
6.30 &
5.30 &
4.60 &
4.55\tabularnewline
$p_{D}=0.99$ &
5.51 &
4.39 &
3.66 &
3.61\tabularnewline
$p_{D}=0.95$ &
5.82 &
4.74 &
4.03 &
3.98\tabularnewline
$p_{D}=0.85$ &
7.07 &
6.17 &
5.52 &
5.48\tabularnewline
\hline 
\end{tabular}
\par\end{centering}
\end{table}

\section{Conclusions\label{sec:Conclusions}}

We have presented the trajectory PHD filter and a Gaussian mixture
implementation. The trajectory PHD filter uses sets of trajectories
as state variable to enable inference over the trajectories, without
the need of evaluating all data association hypotheses. It is based
on propagating a Poisson multitrajectory density through the filtering
recursion and performing a KLD minimisation after each update step.

We have also presented the computationally efficient $L$-scan GMTPHD
filter  for linear/Gaussian models, which can be adapted for nonlinear/non-Gaussian
models using nonlinear Gaussian filters.

\appendices{}

\section{\label{sec:Appendix_KLD}}

In this appendix, we prove Theorem \ref{thm:KLD_minimisation_Poisson}.
A multitrajectory density $\pi\left(\cdot\right)$ can be written
as 
\begin{equation}
\pi\left(\left\{ X_{1},...,X_{n}\right\} \right)=\rho_{\pi}\left(n\right)n!\pi_{n}\left(X_{1:n}\right)
\end{equation}
where $\pi_{n}\left(\cdot\right)$ is a permutation invariant ordered
density with
\begin{align*}
\int\pi_{n}\left(X_{1:n}\right)dX_{1:n} & =1.
\end{align*}
The marginal density of one trajectory of this density is 
\begin{align*}
\tilde{\pi}_{n}\left(X\right) & =\int\pi_{n}\left(X,X_{2:n}\right)dX_{2:n}
\end{align*}
Substituting (\ref{eq:Poisson_prior}) into (\ref{eq:KLD_definition}),
we have that
\begin{align}
\mathrm{D}\left(\pi\left\Vert \nu\right.\right) & =\sum_{n=0}^{\infty}\rho_{\pi}\left(n\right)\log\frac{\rho_{\pi}\left(n\right)}{e^{-\lambda_{\nu}}\lambda_{\nu}^{n}/n!}\nonumber \\
 & \;+\sum_{n=0}^{\infty}\rho_{\pi}\left(n\right)\int\pi_{n}\left(X_{1:n}\right)\log\frac{\pi_{n}\left(X_{1:n}\right)}{\prod_{j=1}^{n}\breve{\nu}\left(X_{j}\right)}dX_{1:n}.\label{eq:append_KLD_min}
\end{align}

We want to find $\lambda_{\nu}$ and $\breve{\nu}\left(\cdot\right)$
that minimise (\ref{eq:append_KLD_min}). By derivating the first
term w.r.t. $\lambda_{\nu}$ and equating it to zero, we obtain that
the unique minimum is achieved by setting $\lambda_{\nu}=\sum_{n=0}^{\infty}n\rho_{\pi}\left(n\right).$
The minimisation over $\breve{\nu}\left(\cdot\right)$ can be done
as in the target case \cite{Angel15_d} , which results in 
\begin{align*}
\breve{\nu}\left(X\right) & =\frac{D_{\pi}\left(X\right)}{\sum_{n=0}^{\infty}\rho_{\pi}\left(n\right)n}
\end{align*}
or, equivalently, $D_{\nu}\left(\cdot\right)=D_{\pi}\left(\cdot\right)$.

\section{\label{sec:Append_TPHD_prediction}}

In this appendix, we prove Theorem \ref{thm:PHD_prediction}. A set
of trajectories at time $k$ can be decomposed as $\mathbf{W}\uplus\mathbf{X}\uplus\mathbf{Y}\uplus\mathbf{Z}$
where $\mathbf{W}$ denotes the set of new born trajectories at time
$k$, $\mathbf{X}$ the set of trajectories present at times $k-1$
and $k$ but not present at $k+1$, $\mathbf{Y}$ the set of trajectories
present at time $k-1$ but not present at time $k$ and $\mathbf{Z}$
the set of trajectories present at a time before $k-1$ but not at
time $k$. We first clarify that if $\left(t,x^{1:i}\right)\in\mathbf{W}$,
then, $t=k$, $i=1$; if it belongs to $\mathbf{X}$, then $t<k$,
$i=k-t+1$; if it belongs to $\mathbf{Y}$, then $t<k$, $i=k-t$;
and finally, if it belongs to $\mathbf{Z}$, then, $t<k-1$, $i<k-t$.
As $\mathbf{W},\mathbf{X},\mathbf{Y}$ and $\mathbf{Z}$ are independent
and Poisson distributed due to Assumptions P2-P3 so we can obtain
their predicted PHDs independently. The overall predicted PHD is then
the sum of these predicted PHDs due to the superposition of Poisson
processes \cite{Angel15_d}. 

We use Theorem 5 in \cite{Angel15_prov}. For dead trajectories, the
prediction step leaves the multitrajectory density unaltered and so
its PHD. The PHD of new born trajectories is analogous to the PHD
of new born targets by setting the time to $k$ and duration to one.
Using Theorem 5 in \cite{Angel15_prov}, we have that for $\mathbf{Y}=\left\{ \left(t_{1},x_{1}^{1:i_{1}}\right),...,\left(t_{n},x_{n}^{1:i_{n}}\right)\right\} $,
\begin{align*}
\omega^{k}\left(\mathbf{Y}\right) & =\pi^{k-1}\left(\left\{ \left(t_{1},x_{1}^{1:i_{1}}\right),...,\left(t_{n},x_{n}^{1:i_{n}}\right)\right\} \right)\\
 & \quad\times\prod_{j=1}^{n}\left(1-p_{S}\left(x_{j}^{i_{j}}\right)\right).
\end{align*}
Using Assumption P3 and (\ref{eq:Poisson_prior}), we get that the
predicted PHD, for $\left(t,x^{1:i}\right)\in\mathbf{Y}$, is 
\begin{align*}
D_{\omega^{k}}\left(t,x^{1:i}\right) & =\left(1-p_{S}\left(x^{i}\right)\right)D_{\pi^{k-1}}\left(t,x^{1:i}\right).
\end{align*}
Similarly, for $\mathbf{X}=\left\{ \left(t_{1},x_{1}^{1:i_{1}}\right),...,\left(t_{n},x_{n}^{1:i_{n}}\right)\right\} $,
\begin{align*}
\omega^{k}\left(\mathbf{X}\right) & =\pi^{k-1}\left(\left\{ \left(t_{1},x_{1}^{1:i_{1}-1}\right),...,\left(t_{n},x_{n}^{1:i_{n}-1}\right)\right\} \right)\\
 & \quad\times\prod_{j=1}^{n}\left(g\left(x_{j}^{i_{j}}\left|x_{j}^{i_{j}-1}\right.\right)p_{S}\left(x_{j}^{i_{j}-1}\right)\right)
\end{align*}
which implies that the predicted PHD for $\mathbf{X}$ is the one
indicated in Theorem \ref{thm:PHD_prediction}, which finishes the
proof. 

\section{\label{sec:Append_TPHD_update}}

In this appendix, we prove Theorem \ref{thm:PHD_update}. As with
the PHD filter, we first compute the density of the measurement \cite{Angel15_d}.
Using (\ref{eq:marginalisation_PHD}) and Assumption U3, the multitarget
predicted density at time $k$ is Poisson with PHD
\begin{align}
D_{\omega_{\tau}^{k}}\left(y\right) & =\sum_{t=1}^{k}\int D_{\omega^{k}}\left(t,x^{1:k-t},y\right)dx^{1:k-t}
\end{align}
where we have used that $\omega^{k}\left(\cdot\right)$ is zero for
trajectories present later than time $k$. Due to the Poisson prior,
the density of the measurement is Poisson with density \cite{Angel15_d}
\begin{align}
\ell^{k}\left(\mathbf{z}^{k}\right) & =e^{-\int p_{D}\left(y\right)l\left(z|y\right)D_{\omega_{\tau}^{k}}\left(y\right)dy-\lambda_{c}}\nonumber \\
 & \,\times\prod_{z\in\mathbf{z}^{k}}\left[\lambda_{c}\breve{c}\left(z\right)+p_{D}\left(y\right)l\left(z|y\right)D_{\omega_{\tau}^{k}}\left(y\right)dy\right].\label{eq:PDF_measurement_Poisson}
\end{align}

Using (\ref{eq:PHD}) and (\ref{eq:update_trajectories}), we calculate
the updated PHD 
\begin{align*}
 & D_{\pi^{k}}(X)\\
 & \quad=\frac{1}{\ell^{k}\left(\mathbf{z}^{k}\right)}\int\ell^{k}\left(\mathbf{z}^{k}\left|\tau^{k}\left(\left\{ X\right\} \cup\mathbf{X}\right)\right.\right)\omega^{k}\left(\left\{ X\right\} \cup\mathbf{X}\right)\delta\mathbf{X}\\
 & \quad=\frac{\lambda_{\omega^{k}}\breve{\omega}^{k}\left(X\right)}{\ell^{k}\left(\mathbf{z}^{k}\right)}\int\ell^{k}\left(\mathbf{z}^{k}\left|\tau^{k}\left(X\right)\cup\tau^{k}\left(\mathbf{X}\right)\right.\right)\omega^{k}\left(\mathbf{X}\right)\delta\mathbf{X}.
\end{align*}
We consider two cases: $X$ is not present at time $k$ and $X$ is
present at time $k$. For $\tau^{k}\left(X\right)=\emptyset$, we
have
\begin{align*}
D_{\pi^{k}}(X) & =\lambda_{\omega^{k}}\breve{\omega}^{k}\left(X\right)=D_{\omega^{k}}(X).
\end{align*}
For $\tau^{k}\left(X\right)\neq\emptyset$, we have that \cite[Eq. (14)]{Angel15_d}
\begin{align*}
 & \ell^{k}\left(\mathbf{z}^{k}\left|\tau^{k}\left(X\right)\cup\tau^{k}\left(\mathbf{X}\right)\right.\right)\\
 & \,=\left(1-p_{D}\left(\tau^{k}\left(X\right)\right)\right)\ell^{k}\left(\mathbf{z}^{k}\left|\tau^{k}\left(\mathbf{X}\right)\right.\right)\\
 & \,+p_{D}\left(\tau^{k}\left(X\right)\right)\sum_{z\in\mathbf{z}^{k}}l\left(z|\tau^{k}\left(X\right)\right)\ell^{k}\left(\mathbf{z}^{k}\setminus\left\{ z\right\} \left|\tau^{k}\left(\mathbf{X}\right)\right.\right)
\end{align*}
where $B\setminus A=\left\{ z\in B|\,z\notin A\right\} $. Using (\ref{eq:PDF_measurement_Poisson})
and following the same steps as in (target) PHD filter derivation
\cite{Angel15_d}, we find
\begin{align*}
D_{\pi^{k}}(X) & =\left(1-p_{D}\left(\tau^{k}\left(X\right)\right)\right)\lambda_{\omega^{k}}\breve{\omega}^{k}\left(X\right)\\
 & \quad+p_{D}\left(\tau^{k}\left(X\right)\right)\lambda_{\omega^{k}}\breve{\omega}^{k}\left(X\right)\\
 & \quad\times\sum_{z\in\mathbf{z}^{k}}\frac{l\left(z|\tau^{k}\left(X\right)\right)}{\lambda_{c}\breve{c}\left(z\right)+\int p_{D}\left(y\right)l\left(z|y\right)D_{\omega_{T}^{k}}\left(y\right)dy},
\end{align*}
which finishes the proof of Theorem \ref{thm:PHD_update}. 

\section{\label{sec:Append_efficient_implementation}}

In this appendix, we derive an $L$-scan single trajectory filter
that jointly updates the density over the last $L$ time steps and
leaves unaltered the density at previous time steps. We use the ADF
so we assume the posterior at time $k$ is of a certain form and then
we perform KLD minimisations to continue with the filtering recursion.
For the sake of notational simplicity, we assume the trajectory exists
at all time steps so we represent a trajectory as $x^{1:k}$. Let
the posterior at time $k$ be
\begin{align}
\pi^{k}\left(x^{1:k}\right) & =p^{k}\left(x^{k-L+1:k}\right)\prod_{i=1}^{k-L}q^{i}\left(x^{i}\right)\label{eq:efficient_imple1}
\end{align}
where $q^{i}\left(\cdot\right)$ is a density for the state at time
step $i<k-L$ and $p^{k}\left(\cdot\right)$ is the joint density
for the last $L$ time steps. That is, the states corresponding to
the last $L$ time steps are considered jointly and the previous states
are independent. 

After the prediction and update on (\ref{eq:efficient_imple1}), we
obtain
\begin{align*}
\pi_{'}^{k+1}\left(x^{1:k+1}\right) & =r^{k+1}\left(x^{k-L+1:k+1}\right)\prod_{i=1}^{k-L}q^{i}\left(x^{i}\right)\\
r^{k+1}\left(x^{k-L+1:k+1}\right) & \propto l\left(z^{k+1}|x^{k+1}\right)g\left(x^{k+1}\left|x^{k}\right.\right)\\
 & \quad\times p^{k}\left(x^{k-L+1:k}\right)
\end{align*}
where $l\left(z^{k+1}|\cdot\right)$ and $g\left(\cdot\left|\cdot\right.\right)$
represent the likelihood and the transition density, respectively.
We obtain the density of the form (\ref{eq:efficient_imple1}) that
minimises the KLD $D\left(\pi_{'}^{k+1}\left\Vert \pi^{k+1}\right.\right)$
with \cite{Bishop_book06} 
\begin{align*}
p^{k+1}\left(x^{k-L+2:k+1}\right) & =\int r^{k+1}\left(x^{k-L+1:k+1}\right)dx^{k-L+1}\\
q^{k+1-L}\left(x^{k+1-L}\right) & =\int r^{k+1}\left(x^{k-L+1:k+1}\right)dx^{k-L+2:k+1}.
\end{align*}

\bibliographystyle{IEEEtran}
\bibliography{4E__Trabajo_Angel_Mis_articulos_Fusion_2018_Trajectory_PHD_filter_Referencias_TPHD_fusion}

\end{document}